\documentstyle[seceq, epsf,wrapfig]{ptptex}

\def\lesssim{\lower.7ex\hbox{${\buildrel < \over \sim}$}}
\def\gtrsim{\lower.7ex\hbox{${\buildrel > \over \sim}$}}

\notypesetlogo

\title{
Atmospheric Neutrinos \footnote{ Talk presented by S. Midorikawa, 
to appaer in the Proceedings of Yukawa International Seminar '95: 
From the Standard Model to Grand Unified Theories.} 
}

\author{
Morihiro {\sc Honda}$^{1}$, Takaaki {\sc Kajita}$^{1}$, 
Katsuaki {\sc Kasahara}$^{2}$, \break and \break
Shoichi {\sc Midorikawa}$^{3}$ 
}

\inst{
$^{1}$ Institute for Cosmic Ray Research, University of Tokyo, Tokyo 188
\\
$^{2}$ Faculty of Engineering, Kanagawa University, Yokohama 221
\\
$^{3}$ Faculty of Engineering, Aomori University, Aomori 030
}

\recdate{
\today
}

\abst{
The atmospheric neutrino-flux is calculated over the wide 
energy range from 1~GeV  to 3,000~GeV for the study of 
neutrino-physics using the data from underground 
neutrino-detectors. 
The uncertainty of atmospheric neutrino--fluxes is also 
discussed.
A brief comment is made to interpret the anomaly 
in terms of neutrino oscillations. 
}

\begin{document}

\maketitle

\section{Introduction}

Since the observation of atmospheric neutrinos by the 
Kamiokande group~\cite{hirata1},
some of the underground detectors~\cite{imb}~\cite{soudan} 
have confirmed that the flux ratio of neutrino species 
${(\nu_\mu + \bar\nu_\mu)/(\nu_e + \bar\nu_e)}$
is significantly different from the expected value, 
although the situation is still controversial
~\cite{frejus}~\cite{nusex}. 
The atmospheric neutrino anomaly may have a crucial importance in 
particle physics, sine it can be interpreted in terms of neutrino 
oscillations with a large mixing angle and a typical mass squared difference 
of ${{\cal O} \left(10^{-2}~\rm eV^2 \right) }$~
\cite{lpw}~\cite{bw}~\cite{hhm}~\cite{hirata2}. 
The observation of the zenith angle variaton of the double ratio
${(\mu/e)_{data}/(\mu/e)_{MC}}$ at multi-GeV energies is also 
suggestive~\cite{fukuda}. It is therefore important to calculate 
atmospheric neutrino fluxes precisely.

Atmospheric neutrino fluxes have been calculated from the incident beam 
of primary cosmic rays by Volkova~\cite{volkova},
Mitsui et al~\cite{mitsui}, Butkevich et al~\cite{butkevich}, and 
Lipari~\cite{lipari} mainly for high energies(from around 1 GeV to above 
100,000 GeV). Gaisser et al\cite{gsb}, Barr et al~\cite{bgs}, 
Bugaev and Naumov~\cite{bn}, Lee ans Koh~\cite{lk}, and  
Honda et al~\cite{hkhm} made a detailed calculation of the 
atmospheric neutrino fluxes for low energies from the primary cosmic rays. 
On the other hand, Perkins~\cite{perkins} calculated the low energy 
atmospheric neutrinos using $\mu$-flux observed at high altitude.  
  
In this paper, we report the detailed calculation of the atmospheric neutrino 
fluxes in the energy range from 30~MeV up to 3,000~GeV, corresponding to the 
observation range of underground neutrino detectors~\cite{hkkm}. 
We also discuss the 
possiblity to interpret the anomaly in terms of neutrino oscillations.

\section {Primary Cosmic Ray Fluxes}

Primary cosmic ray fluxes are relatively well known in the low energy
region($\lesssim$ 100GeV), by which the low energy atmospheric $\nu$-fluxes
($\lesssim$ 3 GeV)are mainly produced.
Webber and Lezniak~\cite{wl} have compiled the energy spectrum of the cosmic 
rays for the hydrogen, helium, and CNO nuclei in the range 10~MeV $\sim$ 
1,000~GeV for three levels of solar activity. 
A similar compilation has been made by others for hydrogen and helium 
nuclei, which agrees well with that of Webber and Lezniak.

\begin{wrapfigure}{l}{8.4cm}
\epsfysize=10cm
\centerline{\epsffile{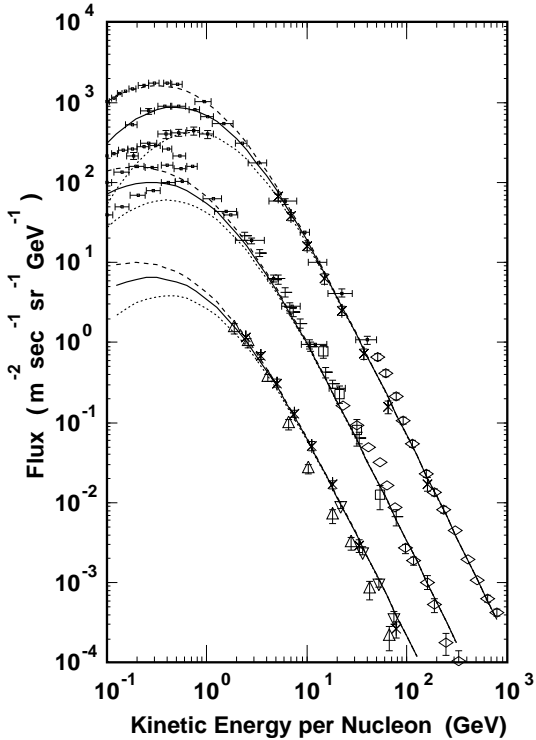}}
\caption{Observed fluxes of cosmic ray protons, helium nuclei, and CNOs
from the compilation of Webber and Leziak$^{1)}$. 
Solid lines are our parametrization for solar mid, dashed lines for 
solar min., and dotted lines for solar max.}
\label{1ry-low}
\end{wrapfigure}

\begin{figure}
\epsfysize=7cm
\centerline{\epsffile{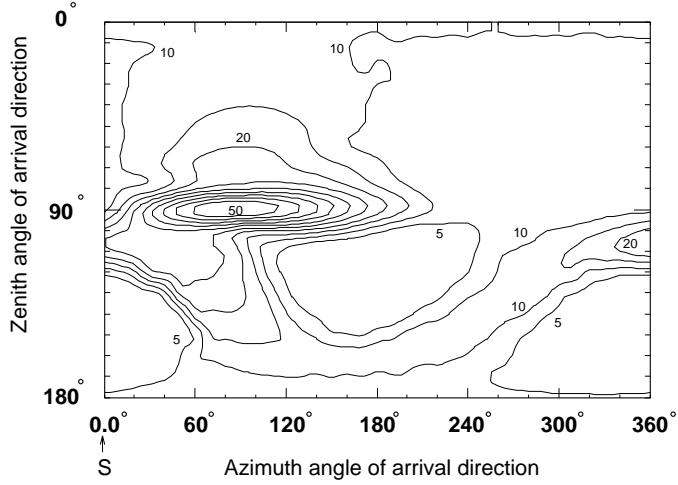}}
\caption{The contour map of cutoff-rigidity for the $\nu$ arrival
directions at Kamioka. Azimuthal angles of $0^\circ, 90^\circ, 
180^\circ$, and $270^\circ$ show directions of south, easth, north, 
and west respectively.}
\label{kamcut}
\end{figure}

The geomagnetic field determines the minimum enegy with which a cosmic 
ray can arrive at the earth. For the cosmic ray nucleus, the minimum 
energy of cosmic rays arriving at the earth is determined by the minimum 
rigidity(rigidity cutoff) rather than the minimum momentum. The value 
of the rigidity cutoff for the actual geomagnetic field can be obtained 
from a computer simulation of cosmic ray trajectories. If a cosmic ray 
particle can reach the earth, the antiparticle with an opposite momentum 
can escape from the earth. We launch antiprotons from the earth, varying 
the position and direction. When a test particle with a given momentum 
reaches a distance of 10 times of the earth's radius, it is assumed that 
the test particle has escaped from the geomagnetic field. The rigidity 
cutoff at Kamioka site is shown as the contour map in Fig. \ref{kamcut}.

\begin{table}
\caption{Compiled cosmic ray spectrum in the form:
${A (E/100~{\rm GeV})^\gamma}$.}
\begin{center}
\begin{tabular}{| c | c | c |} \hline
Nucleus & $A$                              & $\gamma$          \\ \hline
H       & $(6.65 \pm 0.13) \times 10^{-2}$ & $-2.75 \pm 0.020$ \\
He      & $(3.28 \pm 0.05) \times 10^{-3}$ & $-2.64 \pm 0.014$ \\
CNO     & $(1.40 \pm 0.07) \times 10^{-4}$ & $-2.50 \pm 0.06$  \\
Ne-S    & $(3.91 \pm 0.03) \times 10^{-5}$ & $-2.49 \pm 0.04$  \\
Fe      & $(1.27 \pm 0.11) \times 10^{-5}$ & $-2.56 \pm 0.04$  \\ \hline
\end{tabular}
\end{center}
\label{table:fit-parm}
\end{table}

Cosmic rays with energy greater than 100~GeV, which are responsible for 
$\gtrsim 10$~GeV atmospheric neutrino fluxes, are not affected by solar 
activity and by geomagnetic field.  There are few measurements of the 
cosmic ray chemical composition at these energies, especially above 1~TeV. 
We compiled the available data and and parametrized the observed fluxes 
for $\ge$~100 GeV with a single power function, and show the result in 
table \ref{table:fit-parm}. 
We treated bound nucleons at these energies as independent particles, and 
estimated the primary nucleon spectrum.

\section {Production and Decay of hadrons}

As cosmic rays propagate in the atmosphere,
they produce $\pi$'s and $K$'s in interactions with air nuclei.
These mesons create atmospheric $\nu$'s when they decay
as follows:

\begin{equation}
\begin{array}{lllll}
A_{cr} + A_{air} \rightarrow  &\pi^\pm, K^\pm, K^0, \cdots\\
&\pi^+   \rightarrow &\mu^+ + \nu_\mu& & \\
&&\mu^+  \rightarrow &e^+ + \nu_e + \bar \nu_\mu\\
&\pi^-   \rightarrow &\mu^- + \bar \nu_\mu& & \\
&&\mu^-  \rightarrow &e^- + \bar \nu_e + \nu_\mu\\
&.\\
&.\\
\end{array}
\end{equation}

The calculations of the cosmic ray protons with air nuclei consists 
of a number of Monte Carlo codes corresponding to different primary 
energies.
We employed the NUCRIN~\cite{nucrin} Monte Carlo code
for the hadronic interaction of cosmic rays for $E_{lab}\le$ 5~GeV,
and LUND code -- FRITIOF version 1.6~\cite{lund1}
and JETSET version 6.3~\cite{lund2} -- for 
$ 5~{\rm GeV} \le E_{lab} \le 500~{\rm GeV}$.
Above 500~GeV, 
the original code developed by Kasahara et al
(COSMOS)~\cite{kasahara} was used.
The $K/\pi$ ratio
is taken 7~\% at 10~GeV, 11~\% at 100~GeV, and 14~\%
at 1,000~GeV in laboratory energy.

We consider all the decay modes of $\pi$ and $K$ mesons but for 
rare ones.
We have ignored charmed meson production,
since the contribution of charmed particle to atmospheric neutrinos 
becomes sizable only for $E_{\nu} \gtrsim 100$~TeV. 

In the two body decay of charged $\pi$'s and $K$'s, the resulting $\mu^{\pm}$ 
is fully polarized against (toward) the direction of $\mu$ motion 
in the charged $\pi$ or $K$ rest frame.
We took into account the muon polarization effect in the 
subsequent decay following Hayakawa~\cite{hayakawa}.
We applied the discussion in Ref.~28 for the polarization 
of $\mu$'s from the $K_{3l\nu}$ decay.    
The small angle scattering of $\mu$'s in the atmosphere reduces the $\mu$ 
polarization. This depolarization effect was also evaluated in 
Ref.~27 as of the order of 21~MeV/$vp$, where $v$ and $p$ are 
velocity and momentum  of $\mu$'s respectively.

\section {Atmospheric Neutrinos}

\begin{figure}
\epsfysize=8.8cm
\centerline{\epsffile{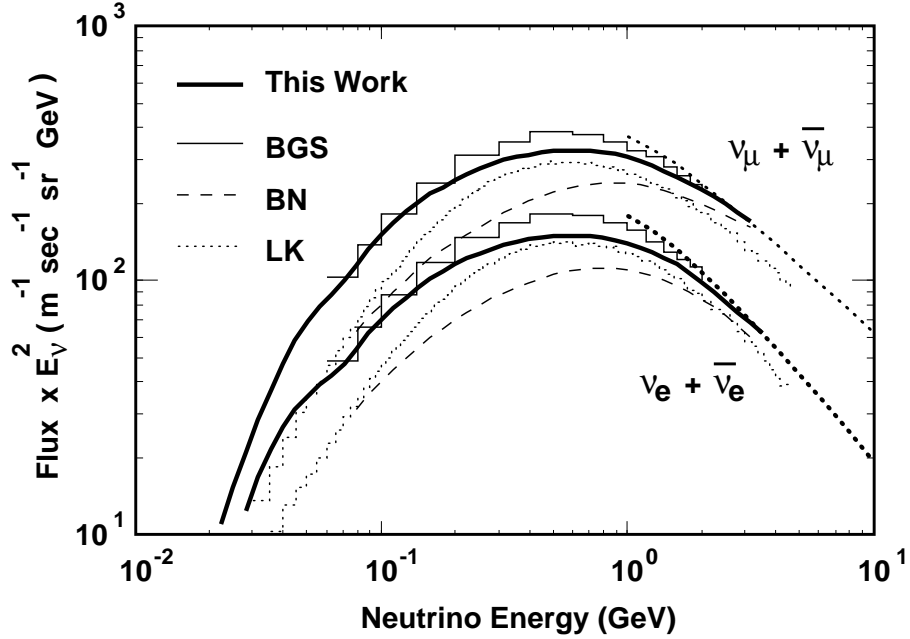}}
\caption{The atmospheric $\nu$--fluxes multiplied by $E_\nu^2$ for
the Kamioka site at solar mid. (solid line).
BGS are from Ref.~[16],
BN from Ref.~[17],
and LK from Ref.~[18].
The dotted line is the result from the calculation for high 
energy without the rigidity cutoff, and averaged over all directions.}
\label{kamnflx}
\end{figure}

At low energies, the rigidity cutoff has a significant directional variation. 
In the one-dimensinoal approximation which we adopted, we expect larger 
$\nu$-fluxes from the low rigidity cutoff directions and a smaller 
$\nu$-fluxes from the high rigidity cutoff direction. In the actual case, 
however, it may be difficult to observe these variations. 
There is a smearing effect of direction in the $\nu$-detector. 
When a low energy $\nu$ (${\lesssim 3}$~GeV) creates a charged lepton 
by a quasi-elastic process, the lepton has a typical angle of 
50 -- 60$^{\circ}$ from the $\nu$ direction. Thus the directional 
dependence of atmospheric neutrino flux is small for lower enregy 
neutrinos, especially when they are observed in the detector. 
We present in Fig.~\ref{kamnflx} the atmospheric neutrino fluxes averaging 
over all directions together with other calculations.    

\begin{figure}
\epsfysize=5.5cm
\centerline{\epsffile{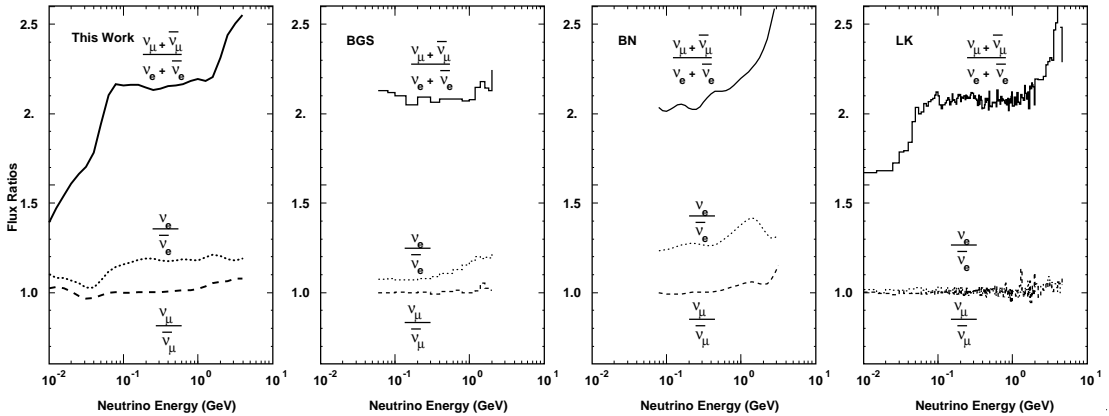}}
\caption{The flux ratio of $\nu$--species calculated for Kamioka.
BGS are from Ref.~[16],
BN from Ref.~[17],
and LK from Ref.~[18]
as before.}
\label{nflx-ratio}
\end{figure}

In Fig. \ref{nflx-ratio}, we show the flux ratio by $\nu$-species along 
with those of other authors. Although the calculation method and some 
of physical assumptions are different among these authors, the ratio 
${(\nu_e + \bar\nu_e)/(\nu_\mu + \bar\nu_\mu)}$ is very similar each 
other. The relatively large difference in ${\bar\nu_e/\nu_e}$ among them 
may reflect the difference of calculation scheme and/or the physical 
assumptions.  

\section{Uncertainties}

The systematic error in the calculation of atmospheric $\nu$--fluxes
comes mainly from the incompleteness of the knowledge of the primary 
cosmic ray fluxes.
Even at low energies, where the primary cosmic ray fluxes are rather well 
studied, it is difficult to determine the absolute value due to the 
uncertainties in the instrumental efficiency ($\sim 12 \%$) and exposure 
factor(2 -- 3\%).  
In our compilation, the error in the fit is $\sim$ 10~\% for the 
nucleon flux at 100~GeV and $\sim$~20~\% at 100~TeV.
Assuming $\sim$~10\% uncertainty below 100~GeV,
the systematic error in the atmospheric $\nu$--fluxes is 
estimated to be $\sim$ 10\% at $\le 3$~GeV, 
increasing to $\sim$~20~\% at 100~GeV, 
and remains almost constant up to 1,000~GeV.

The interaction model is another source of systematic errors.
In our comparison, the agreement of the LUND model and the COSMOS code
with the experimental data is $\lesssim$~10~\%.
The authors of the NUCRIN code claim that the agreement is 
within 10 -- 20~\%~\cite{nucrin}.
The hadronic interaction below 5~GeV contributes 
at most 5~\% to the production of atmospheric $\nu$--fluxes at 1~GeV.
The systematic error caused by the hadronic interaction model is
estimated to be $\sim$ 10~\% above 1~GeV.

One-dimensional approximation which we have adopted is justified only 
at high energies. It is expected to be accurate above 3~GeV. 
Since the calculation of rigidity cutoff is very simplified in this 
scheme, this may result in a systematic error in the absolute value 
of the atmospheric neutrino fluxes of 10 -- 20 \% at 100~MeV and 5 \% 
at 1~GeV.

The statistics of the Monte Carlo calculation also causes an error  
in the atmospheric neutrino fluxes.
The uncertainty due to the statistics is estimated to be 
$\lesssim 5$~\% up to 100 -- 300~GeV
for $\nu_\mu$ and $\bar\nu_\mu$, 
and up to 30 -- 100~GeV for $\nu_e$ and $\bar\nu_e$,
depending on the zenith angle.
The errors increase to $\sim$ 10~\% 
at the highest energy for each kind of $\nu$'s.

Combining all the systematic and non--systematic errors,
the total error is estimated as 
15 \% from 1~GeV to 100~GeV, and 20 -- 25~\% at 
the highest energy in our calculation.
However, the error of the species ratio is smaller than that of 
the absolute value,
since the $\nu$--species ratio is not affected much 
by the uncertainty of primary fluxes and the calculation scheme. 
It is estimated to be
$\lesssim$ 10~\% below 100~GeV for $\nu/\bar\nu$ and
$\lesssim$ 5~\% below 30~GeV for 
$(\nu_\mu + \bar\nu_\mu)/(\nu_e + \bar\nu_e)$.
These errors also increase to 10 -- 15~\% at the highest energies
in our calculation.

\section{Neutrino Oscillations}

We compare the Kamiokande data~\cite{fukuda} with the theotetical 
calculations.
As in our previous paper~\cite{hhm}, we define,

\begin{equation}
<\varepsilon_\alpha \sigma_\alpha F_\beta>=\sum_{\nu, {\bar \nu}} \int
\varepsilon_\alpha(E_\alpha)\sigma_\alpha(E_\nu,E_\alpha)
F_\beta(E_\nu, \Omega_\nu)\rho(h)dE_\alpha dE_\nu d\Omega_\nu dh,
\label{eq1}
\end{equation}

\noindent
where $\varepsilon_\alpha(E_\alpha)$ is the detection efficiency for an
$\alpha$-type charged lepton with energy $E_\alpha$, 
$\sigma_\alpha$ the differencial cross section of $\nu_\alpha$,
$F_\beta(E_\nu, \Omega_\nu)$ the incident $\nu_\beta$ flux with energy 
$E_\nu$ and zenith angle $\Omega_\nu$ .

Insted of the number $N^{obs}_e$ and $N^{obs}_\mu$ 
of the observed electron and muon events, we use the ratios  
defined as follows:

\begin{equation}
f_1 = 
\frac{<\varepsilon_\mu \sigma_\mu F_e>}{<\varepsilon_\mu \sigma_\mu F_\mu>},
\qquad
f_2 =
\frac{<\varepsilon_e \sigma_e F_e>}{<\varepsilon_e \sigma_e F_\mu>}, 
\end{equation}
and 
\begin{eqnarray}
U_e &=& {N^{obs}_e \over {\kappa <\varepsilon_e \sigma_e F_\mu>}} 
= f_2 \frac{N^{obs}_e}{N^{MC}_e} \\
\label{eq2}
U_\mu &=& {N_\mu \over {\kappa <\varepsilon_\mu \sigma_\mu F_\mu>}}  
= \frac{N^{obs}_\mu}{N^{MC}_\mu},
\end{eqnarray}
where ($\kappa = [number~of~nucleons] \times [time]$), 
$N^{MC}_e$ and $N^{MC}_\mu$ are the expected numbers 
of electron and muon events from the Monte Carlo calculations respectively.

If there is no \lq atmospheric neutrino anomaly', the data would 
point to ($f_2$, 1) in the ($U_e, U_\mu$) plain. 
Note that the effects of flux models, geomagnetic cutoffs, and 
the detection efficiencies are all inluded in the values of $f_1$ and $f_2$.
We find  $f_1 \simeq f_2 \simeq 0.473 = f$ from our neutrino fluxes 
and the detection efficiency of the sub-GeV data at Kamiokande.   

If there are neutrino oscillations, the ratios $U_e$ and $U_\mu$ become 

\begin{eqnarray}
U_e   &=& f_2<P(\nu_e \to \nu_e)> + <P(\nu_\mu \to \nu_e)>    \\
U_\mu &=& <P(\nu_\mu \to \nu_\mu)> + f_1<P(\nu_e \to \nu_\mu)>,
\end{eqnarray}

\noindent
where the brackest means the average over the distances and energies of 
neutrinos:   

\begin{eqnarray}
&&<P(\nu_\beta \to \nu_\alpha)>=
\frac{1}{\kappa<\varepsilon_\alpha \sigma_\alpha F_\beta>} \times
\nonumber \\
&&\sum_{\nu, {\bar \nu}} \int
\varepsilon_\alpha(E_\alpha)\sigma_\alpha(E_\nu,E_\alpha)
F_\beta(E_\nu, \Omega_\nu)P(\nu_\beta \to \nu_\alpha)
\rho(h)dE_\alpha dE_\nu d\Omega_\nu dh.
\end{eqnarray}

\begin{table}
\caption{the values of $U_e$, $U_\mu$, and $U_e/U_\mu$ calculated from 
both the sub- and multi-GeV data for the Kamioka site with an exposure 
of 7.7~$ktn \cdot yr$. The value $f$ is taken to be $f = 0.473$.}
\label{U-kam}
\begin{center}
\begin{tabular}{ l | c | c | c } \hline
Energy    &   $U_e$             &   $U_\mu$           &   $U_e/U_\mu$       \\ 
\hline \hline
Sub-GeV   & ${0.518 \pm 0.084}$ & ${0.656 \pm 0.103}$ & ${0.790 \pm 0.072}$ \\
Multi-GeV & ${0.697 \pm 0.126}$ & ${0.832 \pm 0.144}$ & ${0.838 \pm 0.111}$ \\
\hline
\end{tabular}
\end{center}
\end{table}

We summarize in Table \ref{U-kam} the values of $U_e$, $U_\mu$, and 
$U_e/U_\mu$ which are obtained from both the sub- and multi-GeV data 
with 7.7~$ktn \cdot yr$. 
We used only single ring events for the analysis of sub-GeV data. 
For the multi-GeV data, we used both single and multi ring events, 
and evaluated $U_e$ using Eq.~\ref{eq2} with $f_2=f$ obtained 
from the sub-GeV data so that we can compare both data directly. 
We combine the statistical and systematic errors of $\sim 15 \%$ 
including the uncertaity in the neutrino cross section. 

We show in Fig.~5 the regions allowed by the Table~\ref{U-kam} 
together with the region ($U_e, U_\mu$) allowed by three neutrino 
oscillations assuming $f_1 = f_2 \equiv f$.

\begin{figure}
    \begin{center}
	\input{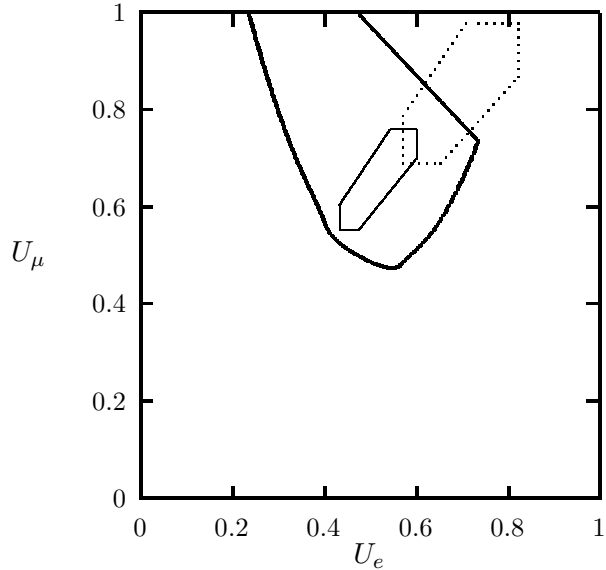}
    \caption{Analysis of the Kamiokande data. 
             Bold line : Allowed region by three neutrino oscllations. 
             Solid line: Kamikande Sub-GeV data.
	     Dashed line: Kamiokande Multi-GeV data.} 
    \end{center}
    \label{ueumu}
\end{figure}

From this figure, we find that the sub-GeV and muti-GeV data are marginally
consistent each other, and that it is possible to explain the anomaly in 
terms of three neutrino oscillations. 
Although it may be premature to deduce what mode dominates oscillatios,
the multi-GeV data seem to sugest 
${\nu_e \longleftrightarrow \nu_\mu}$ oscillations, 
while the sub-GeV data prefer 
${\nu_\mu \longleftrightarrow \nu_\tau}$ oscillations. 
It seems that $U_e$ and $U_\mu$ grow with energy, keeping the $U_\mu/U_e$ 
ratio almost constant. The same conclusion has also been drawn by 
Fogli and Lisi~\cite{fogli}. 
This effect may be explained by the improvement of of the data and/or 
calculations.   
However, if we take both the theory and the experiment seriouly, 
we are lead to a  
more fascinating conjecture.  We can explain both data if matter enhanced 
oscillations would occur between sub- and multi-GeV regions.   

In conclusion, we have shown a detailed calculation of atmospheric 
neutrinos and a new analysis of the anomaly using the recent Kamiokande 
data. The atmospheric neutrino anomaly is eager for further confirmation  
by another types of neutrino expriments such as a long baseline experiment.  

\section*{Acknowledgements}

This work is supported in part by Grant-in-Aid for Scientific Reseach, 
of the Ministry of Education, Science and Culture $\sharp$07640419.

\end{document}